\documentclass[showpacs,preprintnumbers,amsmath,amssymb,12pt,floatfix,epsfig]{revtex4}
\usepackage{dcolumn}
\usepackage{bm}
\usepackage[dvips]{color}
\usepackage{graphicx}

\usepackage{graphicx}
\usepackage{epsfig}

\baselineskip=24ptplus.5ptminus.2pt \vspace*{0.1 in} \large

\def\prl#1{Phys.\ Rev.\ Lett.\ {\bf #1}}
\def\pr#1{Phys.\ Rev.\ {\bf #1}}

\def\pl#1{Phys.\ Lett.\ {\bf #1}}

\def\mo100{${}_{42}^{100}Mo_{58}$}

\def\0nu{0$\nu$}

\def\be{\begin{equation}}
\def\ee{\end{equation}}
\def\Be{\begin{eqnarray}}
\def\Ee{\end{eqnarray}}
\def\ba{\begin{array}}
\def\ea{\end{array}}

\begin{document}
\title{\bf Neutrino reactions via neutral and charged current by
Quasi-particle Random Phase Approximation(QRPA)}
\author{
Myung-Ki Cheoun${^{1)}}$\footnote{Corresponding author :
cheoun@ssu.ac.kr}, Eunja Ha$^{1)}$, K. S.
Kim$^{2)}$\footnote{kyungsik@hau.ac.kr}, Toshitaka
Kajino$^{3,4)}$}
\address{
1) Department of Physics, Soongsil University, Seoul 156-743,
Korea
\\
2) School of Liberal Arts and Science, Korea Aerospace University,
Koyang 412-791, Korea  \\
3) National Astronomical Observatory, Mitaka, Tokyo 181-8589,
Japan \\
4) Department of Astronomy, Graduate School of Science, University
of Tokyo, 7-3-1 Hongo, Tokyo 113-0033, Japan }

\def\ref#1{$^{#1)}$}

\begin{abstract}
We developed the quasi-particle random phase approximation (QRPA)
for the neutrino scattering off even-even nuclei via neutral
current (NC) and charged current (CC). The QRPA has been
successfully applied for the $\beta$ and $\beta \beta$ decay of
relevant nuclei. To describe neutrino scattering, general
multipole transitions by weak interactions with a finite momentum
transfer are calculated for NC and CC reaction with detailed
formalism. Since we consider neutron-proton (np) pairing as well
as neutron-neutron (nn) and proton-proton (pp) pairing
correlations, the nn + pp QRPA and np QRPA are combined in a
framework, which enables to describe both NC and CC reactions in a
consistent way. Numerical results for $\nu - ^{12}$C, $- ^{56}$Fe
and $- ^{56}$Ni reactions are shown to comply with other
theoretical calculations and reproduce well available experimental
data.
\end{abstract}

\pacs{21.60.Jz,23.40.Hc}


\maketitle

\section{Introduction}
Neutrino ($\nu$) (antineutrino (${\bar \nu}$)) scattering with a
complex nucleus plays important roles on understanding the nuclear
structure probed by weak interaction \cite{Kl92,Con72,Suzuki06} as
well as relevant $\nu$ parameters in the $\nu$ physics
\cite{Woosley90,yoshida08}, such as mass hierarchy and mixing
angle $\theta_{13}$ through detailed analysis of nuclear
abundances in the core collapsing supernova (SN) explosion.
Recently, lots of interests have been focused on the $\nu$ process
\cite{Woosley90,yoshida08,Suzuki09,Heg} in the nuclesynthesis
because the emitted neutrino flux is expected to be large enough
to excite the relevant nuclei in spite of the small cross section
due to the weak interaction. Therefore cross sections for the
neutrino(antineutrino)-nucleus ($\nu ({\bar \nu} ) - A$)
scattering with relevant nuclei are to be treated as important
input data for a network calculation estimating the nuclear
abundances, in specific, for the weak rapid process \cite{Wana06}.

Incident $\nu ({\bar \nu})$ energies exploited in the SN explosion
\cite{Woosley90,yoshida08} are focused on the energy range from a
few to tens of MeV because relevant $\nu ({\bar \nu})$ energy
spectra emitted from a proto-neutron star are presumed to be
peaked mostly around the energy region by following a Fermi-Dirac
distribution given by temperature (T) and chemical potential
($\alpha$) \cite{Kolbe03-a}.
\begin{equation} f(E_{\nu}) = { 1 \over {F{ (\alpha)} T^3}}
{  {E_{\nu}^2 } \over { exp [( E_{\nu} / T ) - \alpha] -1 }}
 ~,\end{equation}
where $F{(\alpha)}$ is a nomalization factor. $(T, \alpha)$ can be
chosen for neutrino types. Therefore, the $\nu ({\bar \nu}) - A$
reactions are usually sensitive on the collective motions of
nucleons inside nuclei and proceed via two-step processes, {\it
i.e.} the target nuclei are excited by the incident $\nu ({\bar
\nu}) $ and decayed to the lower energy states with the emission
of some particles \cite{Kolbe03-a}.

The excitation occurs through various transitions, {\it i.e.}
super allowed Fermi ($J^{\pi} = 0^+)$, allowed Gamow Teller (GT)
($J^{\pi} = 1^+)$, spin dipole ($J^{\pi} = 0^- , 1^-, 2^-)$, and
other higher multipole transitions. Therefore dominant
contributions of the two-step process stem from discrete and giant
resonance (GR) states of the compound nucleus. Their typical
excitation energy is below tens of MeV. It should be noticed that
not only isobaric analogue state (IAS) and GT transitions but also
contributions from higher order multipoles could play important
roles in the $\nu -A$ reaction \cite{Suzuki06,Ring08}.

In the experimental side, only a few data measured from
accelerated-based $\nu$ facilities, which are deduced mainly from
$^{12}C$, exist exclusively for the first stage of the two step
processes, that is, the formation of compound nuclei. They have
been reported as $\nu$ flux averaged total cross sections in the
last decade \cite{lsndedar,lsnd,karmenedar,karmencc,Armb98}

\begin{equation} < \sigma_{\nu} > = { {\int d E_{\nu} \sigma_{\nu} (
E_{\nu}) f ( E_{\nu}) } \over { {\int d E_{\nu} f ( E_{\nu}) } }}
~,\end{equation}
where $\sigma_{\nu} ( E_{\nu})$ and $ f ( E_{\nu})$ are $\nu - A$
cross section and corresponding neutrino flux. Unfortunately,
there are no data including the decay processes, that is, the
second stage in the two-step process. Neutrino facilities usually
used neutrino source from the pion decay at rest (DAR) and in
flight (DIF). For the DAR neutrino, one obtains continuous
spectrum for $\nu_e$, which is given as the Michel spectrum
\begin{equation} f(E_{\nu}) = { 96 E_{\nu}^2 \over {m_{\pi}^4}}
( { m_{\mu} - 2 E_{\nu} })
 ~,\end{equation}
and different continuous spectrum for ${\bar \nu}_{\mu}$
\cite{lsndedar}, and mono-energetic ${ \nu}_{\mu}$ fixed as 29.8
MeV. Continuous $\nu_{\mu}$ spectrum is obtained from the DIF
neutrino. Recently, another new method for the production of
intense and pure $\nu$ beams has been proposed, called the
beta-beam \cite{zucch02,volpe}, which is originally planned to
study the CP violation in the lepton sector. The concept of the
beta-beam is designed by using boosted radioactive ions which can
be decayed with emitting neutrinos. For the study of fundamental
interactions and nuclear structure, this new future facility may
produce monochromatic $\nu$ beams through the electron capture of
radioactive ion beam. Of course, these beams depend on the
existence of ions decaying through the electron capture. It was
also suggested how to obtain a mono-energetic $\nu$ beam by using
an electron capture process on a nucleus \cite{sato,bernabeu}.

Most data are focused on $^{12} $C target because the $^{12}$C
nucleus itself serves as a $\nu ({\bar \nu})$ detector. Data for
the inclusive reaction, $^{12}$C$( \nu_e , e^- ) ^{12}$N$^*$, show
about $3.3 \sim 6.7$ in the $10^{-42}$ cm$^2$ unit including
indicated experimental error bars, while data for the exclusive
reaction like $^{12}$C$( \nu_e , e^- ) ^{12}$N$_{g.s.}$ and
$^{12}$C$( \nu , \nu^{'} ) ^{12}$C (15.11 MeV) for the DAR
neutrino are restricted to $6.07 \sim 10.4$ and $8.5 \sim 12.3$ in
the same unit, respectively
\cite{lsndedar,lsnd,karmenedar,karmencc,Armb98}.

The $\nu$ scattering data based on accelerator are mainly
restricted to the formation of compound nuclei. But the excited
states in compound nuclei subsequently decays into other nuclei
with emitting particles such as proton, neutron, alpha, $\gamma$,
and so on \cite{Suzuki06,yoshida08}. To describe the decay
processes at the second stage, one needs additional calculations
for branching ratios into the decay processes like Hauser-Feshbach
(HF) statistical model \cite{Suzuki06,Haus52}. Any final state
interactions (FSI) between outgoing particles and residual nuclei
should be also taken into account at this stage. Of course, for
the nuclear abundance, both two stages are to be successively
considered.

Before presenting our results, we briefly summarize recent
theoretical status about $\nu - A$ scattering with detailed
references. Since the pioneering work by J. S. Cornnell {\it et
al.}\cite{Con72}, which exactly predicted the $\nu- ^{12}$C
reaction via charged current 20 years ago before the advent of the
data, many theoretical
calculations\cite{Suzuki06,Con72,Volpe00,Kolbe95,Ring08,Kolbe03,Ryck02}
have been done and compared to the experimental data, mainly to
the $\nu - ^{12} C$ reaction. Conventional approaches for the $\nu
- A$ scattering in the low energy region are shell model
(SM)\cite{Suzuki06}, random phase approximation
(RPA)\cite{Volpe00}, and Quasi-particle RPA
(QRPA)\cite{Volpe00,Ring08}.

Although results of most SM calculations \cite{Suzuki06,Volpe00}
converge more or less on the experimental data, they inevitably
depend on the particle model space, the given Hamiltonian, the so
called $g_A$ quenching, and so on. The RPA \cite{Volpe00} and the
QRPA \cite{Volpe00,Ring08} calculation usually overestimated the
data by a factor of $4 \sim 5$. More detailed summaries of the
calculations are presented at
Ref.\cite{Kolbe03-a,Volpe00,Kolbe03}. But recent calculation by
proton-neutron QRPA \cite{Ring08} shows more improved results by
simultaneously considering other relevant processes.

In medium heavy or heavy nuclei, neutron-proton (np) pairing as
well as proton-proton (pp) and neutron-neutron (nn) pairing could
be important because of the small energy gap due to the neutron
excess, although its empirical magnitude is small compared to
those of light nuclei \cite{Ch93}. Moreover the np pairing enables
us to consistently describe the CC and NC reactions in a
framework, as shown later on.

On the other hand, the Continuum RPA (CRPA)
\cite{Kolbe95,Kolbe03,Ryck02,Co05-1} includes continuum excitation
spectrum in the quasi-elastic (QE) region in addition to the
discrete and GR states, while most of SM, RPA and QRPA do not take
explicitly the contribution from the QE region into account. The
CRPA shows almost identical results compared to other approaches
for the $\nu - A$ scattering by the DAR neutrino, so that the QE
contribution seems to be small enough to be neglected on a few
tens of MeV region.

In this work, we present our QRPA formalism for the $\nu - A$
reaction via CC and NC type by including nn, pp and np pairing and
compare them with the available experimental data and other
calculations. Our QRPA has successfully described $\beta$, 2$\nu
\beta \beta$ and $0 \nu 2 \beta$ decays \cite{Ch93}. Fairly
reasonable results with the experimental data are obtained. In
Sec.II, we presented our whole formulas for $\nu -A$ scattering,
which is based on the Bardeen Cooper Schriffer (BCS) theory for
the ground state description and the QRPA for the transition to
the excited state. Relevant weak transitions and their operators
are also  summarized in Sec.II with the cross section formula.
Numerical results and discussions are done in Sec.III. Summary and
conclusions are given at Sec.IV.

\section{Formalism}
\subsection{Description of ground state with np pairing}
We start from the following hamiltonian
\begin{equation}
  H  =  H_{0} + H_{int}~,~H_{0}  = {\mathop{\Sigma}_{a^{'} \alpha}} \epsilon_{a \alpha}
c_{a \alpha}^{+} c_{a \alpha}~, H_{int}  =
{\mathop{\Sigma}_{a^{'}b^{'}c^{'}d^{'}, \alpha \beta \gamma
\delta}} V_{a \alpha b \beta c \gamma d \delta}
  c_{a\alpha}^{+}  c_{b\beta}^{+} c_{d \delta}c_{c \gamma}~,
\end{equation}
where the interaction matrix $V$ is the antisymmetrized
interaction with the Baranger hamiltonian \cite{Ba63} in which two
${- 1 \over 2}$ factors, from J and T coupling, are included, so
that the $H_{int}$ in Eq.(4) is equivalent to the usual $ H_{int}
 = {1 \over 4} {\mathop{\Sigma}_{a^{'}b^{'}c^{'}d^{'}, \alpha \beta
\gamma \delta}} {\tilde V}_{a \alpha b \beta c \gamma d \delta}
  c_{a \alpha}^{+}  c_{b \beta}^{+} c_{d \delta}c_{c \gamma}  $.
Roman letters indicate the quantum numbers of the nucleon states,
i.e. $a$ = ($n_{a},l_{a},j_{a}$). But Roman letters with prime
include also the magnetic quantum number i.e. $a^{'}$ =
($n_{a},l_{a},j_{a},m_{a}$). The isospin of real particles is
denoted as a Greek letter while the isospin of quasi-particles is
expressed as a Greek letter with prime. The operator $c^{+}_{a
\alpha}$, $c_{a \alpha}$ stand for the usual creation and
destruction operators of the nucleons in the state of $a^{'}$ with
an isospin $\alpha$. $c_{\bar d \delta} = c_{n_{d},l_{d},j_{d},-
m_{d}} {(-)}^{j_{d} - m_{d}}$ is the time reversed operator of
$c_{d \delta}$.

We transform this hamiltonian by the general HFB transformation to
quasi-particles,
\begin{equation} a_{c \mu^{'}}^{+}  = {\mathop{\Sigma}_{d^{'}
\delta}} (u_{c^{'} \mu^{'} d^{'} \delta} c_{{ d} \delta}^{+} +
v_{c^{'} \mu^{'} d^{'} \delta} c_{{\bar d} \delta})~,~ a_{{\bar c}
\mu^{'}}  = {\mathop{\Sigma}_{d^{'} \delta}} (u_{{\bar c}^{'}
\mu^{'} {\bar d}^{'}
 \delta}^{*}
 c_{{\bar d} \delta}
+ v_{{\bar c}^{'} \mu^{'} {\bar d}^{'} \delta}^{*} c_{{ d}
\delta}^{+})~.
\end{equation}
Here time reversal and spherical symmetries for the nuclei are
presumed, so that we do not mix different single particle angular
momentum states in Eq.(5). Then the Hamiltonian can be represented
in terms of the quasi-particles as follows
\begin{equation}
H^{'}  = H_{0}^{'} + {\mathop{\Sigma}_{a^{'} \alpha^{'}}} E_{a
\alpha^{'} } a_{a \alpha^{'}}^{+} a_{a \alpha^{'}} + H_{qp.int}~.
\end{equation}
Using the transformation of Eq.(5) we obtain the following HFB
equation

\begin{equation}
\left( \begin{array}{cccc} \epsilon_{p}-\lambda_{p} & 0 &
\Delta_{p {\bar p}} & \Delta_{p {\bar n}} \\
0  & \epsilon_{n}-\lambda_{n} & \Delta_{n
{\bar p}} & \Delta_{n {\bar n}} \\
  \Delta_{p {\bar p}} &
 \Delta_{p {\bar n}} & -\epsilon_{p} + \lambda_{p} & 0 \\
  \Delta_{n {\bar p}} &
 \Delta_{n {\bar n}} & 0 & -\epsilon_{n} + \lambda_{n}
  \end{array}\right)_{c}
\left( \begin{array}{c}
u_{\alpha^{'} p} \\ u_{\alpha^{'} n} \\ v_{\alpha^{'} p} \\
v_{\alpha^{'} n} \end{array}\right)_{c}
 =
 E_{c \alpha^{'}}
\left( \begin{array}{c} u_{\alpha^{'} p} \\ u_{\alpha^{'} n} \\
 v_{\alpha^{'} p} \\
v_{\alpha^{'} n} \end{array}\right)_{c} \label{hfbeq}
\end{equation}
, where $E_{c \alpha^{'}}$ is the energy of a quasi particle with
the isospin quantum number ${{\alpha^{'}}}$ in the state $c$. If
we neglect $\Delta_{np}$, this equation reduces to the standard
BCS equation.
 The pairing potentials $\Delta_{p}$, $\Delta_{n}$ and
$\Delta_{pn}$ in Eq.(7) are detailed at Ref \cite{Ch93}.
\subsection{QRPA coupled by neutron-proton pairing}
Excited states, $\vert m; J^{\pi} M \rangle$, in a compound
nucleus, are generated from the ground state of the target
nucleus, which is an even-even nucleus and acts as the BCS vacua
for the quasi-particle, by operating the following one phonon
operator to the initial nucleus
\begin{equation} Q^{+,m}_{JM}  = {\mathop\Sigma_{k l \mu^{'}
\nu^{'} }} [ X^{m}_{(k \mu^{'} l \nu^{'} J)} C^{+}(k \mu^{'} l
\nu^{'} J M)
 - Y^{m}_{(k \mu^{'} l \nu^{'} J)} {\tilde C}(k \mu^{'} l \nu^{'}
  J M)] ~,\end{equation}
where
\begin{equation}
 C(k \mu^{'} l \nu^{'} J M)  =  {\mathop\Sigma_{m_{k} m_{l}}}
C^{JM}_{j_{k} m_{k} j_{l} m_{l}} a_{l \nu^{'}} a_{k \mu^{'}}~,~
{\tilde C}(k \mu^{'} l \nu^{'} J M)  =  (-)^{J-M} C(k \mu^{'} l
\nu^{'} J - M)~
\end{equation}
with a quasi-particle annihilation operator $a_{l \nu^{'}}$ and
Clebsh-Gordan coefficient $C^{JM}_{j_{k} m_{k} j_{l} m_{l}}$. If
neutron-proton pairing is neglected, the phonon operator decouples
to two phonon operators. One is for the charge changing reaction
such as beta decay and CC neutrino reaction. The second is for the
charge conserving such as electro-magnetic and NC neutrino
reactions. The amplitudes $X_{a \alpha, b \beta}$ and $Y_{a
\alpha, b \beta}$, which stand for forward and backward going
amplitudes from state ${a \alpha}$ to ${b \beta}$, are obtained
from the following QRPA equation, whose detailed derivation was
also shown at Ref. \cite{Ch93}
\begin{equation}
\left( \begin{array}{cccccc}
 ~A_{klab}^{1111} & ~A_{klab}^{1122} &
~A_{klab}^{1112} &
           ~B_{klab}^{1111} & ~B_{klab}^{1122} &  ~B_{klab}^{1112} \\
           ~A_{klab}^{2211} & ~A_{klab}^{2222} &  ~A_{klab}^{2212} &
           ~B_{klab}^{2211} & ~B_{klab}^{2222} &  ~B_{klab}^{2212} \\
           ~A_{klab}^{1211} & ~A_{klab}^{1222} &  ~A_{klab}^{1212} &
           ~B_{klab}^{1211} & ~B_{klab}^{1222} &  ~B_{klab}^{1212} \\
                            &                  &            &
                            &                  &            \\
           - B_{klab}^{1111} & -B_{klab}^{1122} &  -B_{klab}^{1112} &
           - A_{klab}^{1111} & -A_{klab}^{1122} &  -A_{klab}^{1112} \\
           - B_{klab}^{2211} & -B_{klab}^{2222} &  -B_{klab}^{2212} &
           - A_{klab}^{2211} & -A_{klab}^{2222} &  -A_{klab}^{2212} \\
           - B_{klab}^{1211} & -B_{klab}^{1222} &  -B_{klab}^{1212} &
           - A_{klab}^{1211} & -A_{klab}^{1222} &  -A_{klab}^{1212} \end{array} \right)
\left( \begin{array}{c}   {\tilde X}_{(a1b1)J}^{m}  \\ {\tilde X}_{(a2b2)J}^{m} \\
  {\tilde X}_{(a1b2)J}^{m} \\
     \cr {\tilde Y}_{(a1b1)J}^{m} \\ {\tilde Y}_{(a2b2)J}^{m} \\
     {\tilde Y}_{(a1b2)J}^{m}  \end{array} \right)
 = \hbar {\Omega}^{m}
 \left ( \begin{array}{c} {\tilde X}_{(k1l1)J}^{m}  \\
{\tilde X}_{(k2l2)J}^{m} \\
 {\tilde X}_{(k1l2)J}^{m} \\  \\
{\tilde Y}_{(k1l1)J}^{m} \\ {\tilde Y}_{(k2l2)J}^{m} \\ {\tilde
Y}_{(k1l2)J}^{m}  \end{array} \right) ~.\label{qrpaeq}
\end{equation}
The A and B matrices are given by
\begin{eqnarray}
A_{klab}^{\gamma^{'} \delta^{'} \mu^{'} \nu^{'}} & = &(E_{k
\gamma^{'}} + E_{l \delta^{'}}) \delta_{ka} \delta_{\gamma^{'}
\mu^{'}} \delta_{l b} \delta_{\delta^{'} \nu^{'}} \\ \nonumber & &
- \sigma_{k \gamma^{'} l \delta^{'}} \sigma_{a \mu^{'} b \nu^{'}}
\mathop{\Sigma}_{\alpha \beta \gamma \delta} [G(k \alpha l \beta a
\gamma b \delta J)(u_{k \gamma^{'} \alpha} u_{l \delta^{'} \beta}
                                     u_{a \mu^{'} \gamma} u_{b \nu^{'} \delta}
                                +     v_{k \gamma^{'}
                                 \alpha} v_{l \delta^{'} \beta}
                                     v_{a \mu^{'} \gamma} v_{b \nu^{'} \delta})
\\ \nonumber  & & + F(k \alpha l \beta a  \gamma b \delta J)(u_{k \gamma^{'}
\alpha} v_{l \delta^{'} \beta}
                                     u_{a \mu^{'} \gamma} v_{b \nu^{'} \delta}
                                +     v_{k \gamma^{'}
                                 \alpha} u_{l \delta^{'} \beta}
                                     v_{a \mu^{'} \gamma} u_{b \nu^{'} \delta})
\\ \nonumber & & -(-1)^{j_{k^{'}}+j_{l^{'}}+J} F(k \alpha l \beta b \delta  a
\gamma J)(u_{k \gamma^{'} \alpha} v_{l \delta^{'}\beta}
                                     u_{b \nu^{'} \gamma} v_{a \mu^{'} \delta}
                                +     v_{k \gamma^{'} \alpha}
                                 u_{l \delta^{'} \beta}
                                     v_{b \nu^{'} \gamma} u_{a \mu^{'}
                                     \delta})~
                                    ]~,
\end{eqnarray}
\begin{eqnarray}
B_{klab}^{\gamma^{'} \delta^{'} \mu^{'} \nu^{'}} & = & - \sigma_{k
\gamma^{'} l \delta^{'}} \sigma_{a \mu^{'} b \nu^{'}}
\mathop{\Sigma}_{\alpha \beta \gamma \delta} [-~G(k \alpha l \beta
a  \gamma b \delta J)(u_{k \gamma^{'} \alpha} u_{l \delta^{'}
\beta}
                                     v_{a \mu^{'} \gamma} v_{b \nu^{'} \delta}
                                +     v_{k \gamma^{'} \alpha} v_{l
                                \delta^{'} \beta}
                                     u_{a \mu^{'} \gamma} u_{b \nu^{'} \delta})
\\ \nonumber & & +~ F(k \alpha l \beta a  \gamma b \delta J)(u_{k \gamma^{'}
\alpha} v_{l \delta^{'} \beta}
                                     v_{a \mu^{'} \gamma} u_{b \nu^{'} \delta}
                                +     v_{k \gamma^{'} \alpha} u_{l
                                \delta^{'} \beta}
                                     u_{a \mu^{'} \gamma} v_{b \nu^{'} \delta})
\\ \nonumber  & & -~(-1)^{j_{k^{'}}+j_{l^{'}}+J} F(k \alpha l \beta b \delta
a \gamma J)(u_{k \gamma^{'} \alpha} v_{l \delta^{'} \beta}
                                     u_{a \mu^{'} \gamma} v_{b \nu^{'} \delta}
                                +     v_{k \gamma^{'} \alpha} u_{l
                                \delta^{'} \beta}
                                     v_{a \mu^{'} \gamma} u_{b \nu^{'} \delta})
] ~,\end{eqnarray}
where $u$ and $v$ coefficients related to the occupation
probabilities of a given state are determined from the HFB
calculation in Eq.(7) with the pairing strength $g_{nn} , g_{pp} $
and $g_{np} $ adjusted to the empirical pairing gaps $\Delta_{nn}
, \Delta_{pp} $ and $\delta_{np}$, respectively \cite{Ch93}. $E_{k
\gamma^{'}}$ indicates the quasi-particle energy of the state $k$
with the quasi-particle isospin $\gamma^{'}$, and $\sigma_{k
\gamma^{'} l \delta^{'}}$ = 1 if $k = l$ and $\gamma^{'}$ =
$\delta^{'}$, otherwise $\sigma_{k \gamma^{'} l \delta^{'} }$ =
$\sqrt 2$. The $G (F) $ matrices are two body particle - particle
(hole) matrix elements obtained as solutions of the following
Bethe - Goldstone equation, called as ${\cal G}$ matrix,
\begin{equation} {\cal G}(w)_{ab,cd}  = V_{ab,cd}^{OBEP} +
 V_{ab,cd}^{OBEP} { Q_{p} \over {w
- H_{0}}} {\cal G}(w)_{ab,cd}~, \end{equation}
where $a, b, c,d$ indicate the single nucleon basis states
characterized by oscillator type wave functions with single
particle energies from the Woods-Saxon potential. $H_{0}$ is the
harmonic oscillator hamiltonian and $Q_{p}$ is the Pauli operator.
$V_{ab,cd}^{OBEP}$ is the phenomenological nucleon-nucleon
potential. We used the one boson exchange potential of the Bonn
group \cite{Ho81}. Of course, one can use effective interactions
for the 2-body interactions, such as the effective Skyrme force
and the surface delta interactions \cite{Ring08} instead of the
${\cal G}$-matrix.

\subsection{Description of CC and NC reaction}
Under the second quantization, matrix elements of any transition
operator ${\cal {\hat O}}$ between a ground state and an excited
state $ | \omega ; J M >$ can be factored as follows
\begin{equation} < QRPA || {\cal {\hat O}}_{\lambda } || ~ \omega ; JM
>  =  {[\lambda]}^{-1} {\mathop{\Sigma}_{ab}} < a ||  {\cal {\hat O}}_{\lambda} || b>
<  QRPA || {[c_a^+ {\tilde c}_b]}_{\lambda} || \omega ; J M > ~.
\end{equation}
Here the first factor $< a ||{\cal {\hat O}}_{\lambda} || b
>$ can be calculated model independently for a given single
particle basis \cite{Don79}. Ground and excited states developed
in the previous subsection are exploited for the second factor
with the Quasi Boson Approximation (QBA). By using the phonon
operator $Q^{+,m}_{JM}$ in Eq.(8), we obtain the following
expression for NC and CC neutrino reactions. For NC reaction,
\begin{eqnarray}
& & <  QRPA || {\cal {\hat O}}_{\lambda } || ~ \omega ; JM
>  \\ \nonumber
= & & {\mathop\Sigma_{a \alpha^{'} b \beta^{'}}}  [ {\cal N}_{a
\alpha^{'} b \beta^{'} } < a \alpha^{'} || {\cal {\hat
O}}_{\lambda}  || b \beta^{'}
>  ~[ u_{pa \alpha^{'}} v_{pb
\beta^{'}} X_{a \alpha^{'} b \beta^{'}} +
v_{pa \alpha^{'}} u_{pb \beta^{'}} Y_{a \alpha^{'} b \beta^{'}} ] \\
\nonumber & & - {(-)}^{j_a + j_b + J } {\cal N}_{ b \beta^{'} a
\alpha^{'} } <b \beta^{'}|| {\cal {\hat O}}_{\lambda}  || a
\alpha^{'}  >~ [ u_{pb \beta^{'}} v_{pa \alpha^{'}} X_{a
\alpha^{'} b \beta^{'}} +v_{pb \beta^{'}} u_{pa \alpha^{'}}  Y_{a
\alpha^{'} b \beta^{'}} ]] + (p \rightarrow n)~,
\end{eqnarray}
where the nomalization factor is given as $ {\cal N}_{a \alpha^{'}
b \beta^{'}} (J) = {\sqrt{ 1 - \delta_{ab} \delta_{\alpha^{'}
\beta^{'} } (-1)^{J + T} }}/ ({1 + \delta_{ab}\delta_{\alpha^{'}
\beta^{'} } }) $. Without the np pairing correlation, this
expression can be reduced to the following simple form
\begin{eqnarray}
& &< QRPA || {\cal {\hat O}}_{\lambda } || ~ \omega ; JM
>  \\ \nonumber
= & & {\mathop\Sigma_{a b}}  [ {\cal N}_{a p b p } < a p || {\cal
{\hat O}}_{\lambda}  || b p> ~[ u_{pa} v_{pb} X_{a p b p} +
v_{pa } u_{pb } Y_{a p b p} ] \\
\nonumber & & - {(-)}^{j_a + j_b + J } {\cal N}_{ b p a p } <b p||
{\cal {\hat O}}_{\lambda}  || a p  > ~[  u_{pb } v_{pa } X_{a p b
p} + v_{pb} u_{pa }  Y_{a p b p} ]] + (p \rightarrow n)~,
\end{eqnarray}
where summations for $\alpha^{'}$ and $\beta^{'}$ are performed in
Eq.(15) since quasi particle $\alpha^{'}$ and $\beta^{'}$ (= 1,2)
include the np pairing. On the other hand, for CC reaction, they
are given as
\begin{eqnarray}
& &< QRPA || {\cal {\hat O}}_{\lambda } || ~ \omega ; JM
>  \\ \nonumber
= & & {\mathop\Sigma_{a \alpha^{'} b \beta^{'}}}  [ {\cal N}_{a
\alpha^{'} b \beta^{'} } < a \alpha^{'} || {\cal {\hat
O}}_{\lambda}  || b \beta^{'}
>  ~[ u_{pa \alpha^{'}} v_{nb
\beta^{'}} X_{a \alpha^{'} b \beta^{'}} + v_{pa \alpha^{'}} u_{nb
\beta^{'}} Y_{a \alpha^{'} b \beta^{'}} ]~.
\end{eqnarray}
This form is also easily reduced to the results by pnQRPA without
pn pairing
\begin{equation} < QRPA || {\cal {\hat O}}_{\lambda } || ~ \omega ; JM
>  = {\mathop\Sigma_{ap bn}}  [ {\cal N}_{a p b n } < a p || {\cal
{\hat O}}_{\lambda}  || b n> ~[ u_{pa} v_{nb} X_{a p b n} + v_{pa
} u_{nb } Y_{a p b n} ]~.
\end{equation}
Aa a simple application of the above form, we can calculate
${\beta}^{\pm}$ decay, whose operator is defined as
\begin{equation} \beta_{LM}^{-}  = {\hat L}^{-1} < p \vert \vert
{\hat \beta} \vert \vert n > [ c_p^+ {\tilde c}_n ]_{LM}~,~
\beta_{LM}^{+}  =  {( \beta_{LM}^{-} )}^+  = {(-)}^{M}
\beta_{L,-M}^{-}~, \end{equation}
as follows in our QRPA formalism \cite{Ch93},
\begin{eqnarray}
 {\langle 1_{m}^{+} \vert \vert
\tau^{+} {\vec \sigma} \vert \vert 0_{i}^{+} \rangle} & = &
{\mathop\Sigma_{ab \alpha^{'} \beta^{'}}} [ X_{(a \alpha^{'} b
\beta^{'}
 )1}^{m}u_{a \alpha^{'} p}v_{b \beta^{'} n} +
Y_{(a \alpha^{'} b \beta^{'} )1}^{m} v_{a \alpha^{'} p}u_{b
\beta^{'} n}] Coef(a,b)~, \\ \nonumber {  \langle 0_{f}^{+} \vert
\vert \tau^{+} {\vec \sigma} \vert \vert 1_{m^{'}}^{+} \rangle} &
= &{\mathop\Sigma_{ab \alpha^{'} \beta^{'}}} [ {\bar X}_{( b
\beta^{'} a \alpha^{'} )1}^{m^{'}}{\bar v}_{a \alpha^{'} p}{\bar
u}_{b \beta^{'} n} + {\bar Y}_{( b \beta^{'} a
\alpha^{'})1}^{m^{'}} {\bar u}_{a \alpha^{'} p}{\bar v}_{b
\beta^{'} n}] Coef(b,a) \end{eqnarray} with
\begin{equation} Coef(a,b)  =~2~\langle (l_{a} {1 \over 2})
j_{a} \vert \vert  \sigma
   \vert \vert (l_{b} {1 \over 2} ) j_{b} \rangle ~.\end{equation}

\subsection{Neutral and Charged Current Operators}

For nuclear weak current operators, we start from a weak current
on the nucleon level. The weak current operator $W^{\mu}$ takes a
$V^{\mu} - A^{\mu}$ current form by the standard electro-weak
theory, which has isoscalar and isovector parts for the NC
interaction \cite{Ch93}
\begin{eqnarray}
W^{\mu} & = & V_{3}^{\mu} - A_{3}^{\mu} - 2 {\sin}^2
{\theta}_{W} J_{em}^{\mu} - { 1 \over 2} ( V_s^{\mu} - A_s^{\mu}) \\
\nonumber & = & ( 1 - 2 {\sin}^2 {\theta}_{W} ) V_{3}^{\mu} -
A_{3}^{\mu} - 2 {\sin}^2 {\theta}_{W} V_{0}^{\mu} - { 1 \over 2} (
V_s^{\mu} - A_s^{\mu}) ~,
\end{eqnarray}
with Weinberg angle $\theta_W$, where we used $J_{em}^{\mu} =
V_3^{\mu} + V_0^{\mu}$. $V_3^{\mu} (= V_{ 1 + i 2}^{\mu})$ and
$A_3^{\mu} (= A_{ 1 + i 2}^{\mu})$ are plus components of the
isovector $V_{i}^{\mu}$ and $A_{i}^{\mu}$ by isospin rotation.
Strangeness contributions, which are isoscalar parts, could be
considered in $ - { 1 \over 2} (V_{s}^{\mu} -  A_{s}^{\mu})$.

For the CC interaction, only $V_{3}^{\mu} - A_{3}^{\mu}$ term is
involved, while $J^{\mu}_{em} = V_{3}^{\mu} + V_{0}^{\mu}$ is
concerned with meson electro-production. Therefore the CC reaction
of the $\nu ({\bar \nu})$ scattering is nearly independent of the
strangeness contents. For the elastic scattering of polarized
electron on the nucleon, $J^{\mu} = - 2 {\sin}^2 {\theta}_{W}
J_{em}^{\mu} - { 1 \over 2} V_s^{\mu} $ is exploited.

For a free nucleon, the current operator comprises the vector, the
axial vector and the pseudo scalar form factor, $F_i^V (Q^2)$,
$F_A (Q^2)$ and $F_P (Q^2)$
\begin{equation}
{W}^{\mu}=F_{1}^V (Q^2){\gamma}^{\mu}+ F_{2}^V (Q^2){\frac {i}
{2M_N}}{\sigma}^{\mu\nu}q_{\nu} + F_A(Q^2) \gamma^{\mu} \gamma^5 +
{ F_P(Q^2) \over {2M}} q^{\mu} \gamma^5~,
\end{equation}
where we take the scalar and the tensor form factor to be zero
because of current conservation and no existence of second class
current, respectively. By the conservation of the vector current
(CVC) hypothesis with the inclusion of the isoscalar strange quark
contributions $F_i^s (Q^2)$, and the vector form factors for
protons and neutrons $F_{i}^{V,~p(n)} (Q^2)$ are expressed as
\cite{giusti1}
\begin{eqnarray}
F_i^{V,p(n)} ( Q^2) &=&({\frac 1 2} - 2 \sin^2 \theta_W )
F_i^{p(n)} ( Q^2) - {\frac 1 2} F_i^{n(p)}( Q^2) -{\frac 1 2}
F_i^s ( Q^2)~~ \mbox{for~ NC}\\ \nonumber & =& ( F_i^{p} ( Q^2) -
 F_i^{n}( Q^2))~~~~~~~~~~~~~~~~~~~~~~~~~~~~~~~~~~~~~~~~~~\mbox{for~ CC}~.
\end{eqnarray}
The axial form factor is usually given by \cite{musolf}
\begin{eqnarray}
F^{NC}_A (Q^2) &=&{\frac 1 2} (\mp g_A + g_A^s)/(1+Q^2/M_A^2)^2~~
\mbox{for~ NC}~\\ \nonumber F_A^{CC} (Q^2) & =& - g_A / {( 1 + Q^2
/ M_A^2)}^2~~~~~~~~~~~~~ \mbox{for~ CC}~, \label{gs}
\end{eqnarray}
where $g_A$ and $M_A$ are the axial coupling constant and the
axial cut off mass, respectively. $-(+)$ coming from the isospin
dependence denotes the knocked-out proton (neutron), respectively.
Since the energy region considered here is below the quasi-elastic
region, the strangeness contributions are not taken into account
in this report. Since we take $+$ sign for $F_A (Q^2) $ in
Eq.(23), the axial form factor in Eq.(25) is just negative to the
form factor elsewhere, for example, in Ref.\cite{giusti1}.

\subsection{Relevant Operators for weak interactions}
In order to calculate weak interactions with nuclei, one resorts
to the Hamiltonian, $ {\cal H} ({\bf x}) = - { { G_F } \over {
{\sqrt 2} }} \int d {\bf x} {\hat W}_{\mu}^{weak} ( {\bf x})
A_{\mu} (q {\bf x}) $, contracted with weak hadron (nuclear)
current operator, ${\hat W}_{\mu}^{weak} ( {\bf x}) $ and external
weak field, $A_{\mu} ( q {\bf x}) = {l}_{\mu} exp( - i {\bf q}
\cdot {\bf x})$ with $q = | {\bf q}|$ and lepton current
${l}_{\mu}$ \cite{Con72}. Here we follow the kinematics at Ref.
\cite{Wal75}. Four momentum transfer is defined as $q^2 (=
q_{\mu}^2 = {(k_{\mu} - \nu_{\mu})}^2 ) = q_0^2 - {\bf q}^2 = -
Q^2 \le 0 $, where $ k_{\mu}$ and ${\nu}_{\mu} $ are final and
initial lepton momenta. The excitation energy of a compound
nucleus is given as $\omega = - q_0 \ge 0 $.

The weak field can be expanded in terms of multipole operators by
using two basic operators
\be { M}_J^{M_J} ( q {\bf x})  =  j_J ( q {\bf x}) Y_J^{M_J} (
\Omega_x )~, ~  {\bf M}_{J L}^{M_J} ( q {\bf x})  = j_J ( q {\bf
x}) {\bf Y}_{J L 1}^{M_J} ( \Omega_x )~, \ee
where vectorial spherical harmonic ${\bf Y}_{J L 1}^{M_J} (
\Omega_x )$ is expressed in term of spherical harmonic $Y_{L}^{m}
( \Omega_x )$, {\it i.e.} ${\bf Y}_{J L 1}^{M_J} ( \Omega_x ) =
 {\mathop\Sigma_{m \lambda}} < L m 1
\lambda \vert ( L 1 ) J M_J
> Y_{L}^{m} ( \Omega_x ) {\bf e}_{\lambda} $ with the spherical unit tensor ${\bf e}_{\lambda}$.
If we make non-relativistic reduction of the one body nucleon
operator using Dirac wave function, we reexpress the one-body weak
transition current in terms of 4 different transition operators
(Coulomb, longitudinal, electric and magnetic) as follows
\Be {\cal { {\hat M}}}_{JM;T M_T} (q {\bf x}) & = &  [ F_1^{(T)}
M_J^{M_J} ( q {\bf x} ) - i { q \over M} [ { F_A^{(T)}
{{\Omega}}_J^{M_J} ( q {\bf x} ) + {{ F_A - \omega F_P^{(T)}}
\over 2} {\Sigma^{''}}_{J}^{M_J} ( q {\bf x}) } ]] I_T^{M_T},
\\ \nonumber
~{\cal {{\hat L}}}_{JM;T M_T} (q {\bf x}) & = & [{- {\omega  }
\over { q } } F_1^{(T)} M_J^{M_J} ( q {\bf x} ) + i ( F_A^{(T)} -
{{q^2  }\over {2 M_N } } F_P^{(T)} ) {\Sigma^{''}}_{J}^{M_{J}} ( q
{\bf x} ) ]I_T^{M_T}~,\\
\nonumber {\cal { {\hat T}}}_{JM;T M_T}^{el} (q {\bf x}) & = & [{
q \over M} [ { F_1^{(T)} {{\Delta}^{'}}_J^{M_J} ( q {\bf x} ) + {1
\over 2} \mu^{(T) } {\Sigma}_{J}^{M_J} ( q {\bf x}) }] + i
F_A^{(T)} {\Sigma^{'}}_{J}^{M_{J}} ( q {\bf x} ) ] I_T^{M_T}~,
\\ \nonumber
{\cal { {\hat T}}}_{JM;T M_T}^{mag} (q {\bf x}) & = & -i { q \over
M} [[ { F_1^{(T)} {{\Delta}}_J^{M_J} ( q {\bf x} ) - {1\over 2}
\mu^{(T) } {\Sigma^{'}}_{J}^{M_J} ( q {\bf x}) }] +   F_A^{(T)}
{\Sigma}_{J}^{M_{J}} ( q {\bf x} ) ] I_T^{M_T}~,
 \Ee
where the superscript $T ( = 0,1) $ means isoscalar and isovector.
The 8 relevant single particle operators ($ M_J^{M_J}
,{{\Omega}}_J^{M_J}, {{\Omega^{'}}}_J^{M_J}, {\Sigma}_{J}^{M_J},
{{\Delta}}_J^{M_J}, {\Sigma^{'}}_{J}^{M_J},
{\Sigma^{''}}_{J}^{M_J}, {{\Delta}^{'}}_J^{M_J} $) are given as
follows, whose detailed derivation was done at Ref.
\cite{Con72,Don79},
\Be M_{J}^{M}~, & & \\ \nonumber \Delta_{J}^{M_{J}} & = &
{\bf M}_{JJ}^{M_J} ( q {\bf x}) \cdot { 1 \over q} \nabla~,\\
\nonumber {\Delta^{'}}_{J}^{M_{J}} & =& {[J]}^{-1} [- J^{1 \over
2} {\bf M}_{JJ+1}^{M_J} ( q {\bf x}) + {( J+ 1)}^{1/2} {\bf
M}_{JJ-1}^{M_J}  ( q {\bf x}) ] \cdot { 1 \over q} \nabla \\
\nonumber \Sigma_{J}^{M_{J}} & = & {\bf M}_{JJ}^{M_J} ( q {\bf x})
\cdot {\sigma} ~~,~~{\Sigma^{'}}_{J}^{M_{J}}  = {[J]}^{-1} [- J^{1
\over 2} {\bf M}_{JJ+1}^{M_J} ( q {\bf x}) + {( J+ 1)}^{1/2} {\bf
M}_{JJ-1}^{M_J}  ( q {\bf x}) ] \cdot \sigma
\\ \nonumber
{\Sigma^{"}}_{J}^{M_{J}} & = & {[J]}^{-1} [ {(J+1)}^{1 \over 2}
{\bf M}_{JJ+1}^{M_J} ( q {\bf x}) + { J}^{1/2} {\bf
M}_{JJ-1}^{M_J} ( q {\bf x}) ] \cdot \sigma
\\ \nonumber
\Omega_J^{M_J} ( q {\bf x}) & = & M_J^{M_J} ( q {\bf x} ) {\bf
\sigma} \cdot { 1 \over q} \nabla ~~,~~ {\Omega^{'}}_J^{M_J} ( q
{\bf x}) = {\Omega}_J^{M_J} ( q {\bf x}) + { 1/2}
{\Sigma^{"}}_J^{M_J} ( q {\bf x}) ~.\Ee
In actual calculation, the last operator is not used because it
can be combined by other operators \cite{Ring08}. Single nucleon
form factors in Eq.(27) are denoted as $F_X^{(T)} (Q^2) $ with $T
= 0,1$ and $X = 1,2,A,P$ standing for Dirac ($X$=1), Pauli
($X$=2), axial, and pseudo-scalar form factors, respectively.
Detailed form factors are referred from Ref. \cite{Con72,Don79}.
The induced pseudoscalar form factor is usually parameterized by
the Goldberger-Treimann relation
\begin{equation}
F_P(Q^2) = {\frac {2M_N} {Q^2+m^2_{\pi}}} F_A(Q^2),
\end{equation}
where $m_{\pi}$ is the pion mass. The contribution of the
pseudoscalar form factor vanishes for the NC reaction because of
the negligible final lepton mass participating in this reaction.
But it can contribute to the CC reaction, in particular, for
$\nu_{\mu} - A$ reaction.

\subsection{Cross Sections}
Based on the initial and final nuclear states, cross section for
$\nu ({\bar \nu})-A$ scattering through the relevant transition
operator in Eq.(27) is given as \cite{Wal75}
\Be & & ({{d \sigma_{\nu}} \over {d \Omega }  })_{(\nu / {\bar
\nu})} = { { G_F^2 \epsilon k } \over {\pi ~ (2 J_i + 1 ) }}~
\bigl[ ~ {\mathop\Sigma_{J = 0}} (
 1+ {\vec \nu} \cdot {\vec \beta }){| <  J_f || {\cal {\hat M}}_J || J_i > | }^2
 \\ \nonumber & & + (
 1 - {\vec \nu} \cdot {\vec \beta } + 2({\hat \nu} \cdot {\hat q} )
 ({\hat q} \cdot {\vec \beta}  ))
  {| <  J_f || {\cal {\hat L}}_J ||
J_i > | }^2  -
 {\hat q} \cdot ({\hat \nu}+ {\vec \beta} )  { 2 Re < J_f || {\cal {\hat L}}_J  || J_i>
{< J_f|| {\cal {\hat M}}_J || J_i >}^*  } \\
\nonumber & &  + {\mathop\Sigma_{J = 1}} ( 1 - ({\hat \nu} \cdot
{\hat q} )({\hat q} \cdot {\vec \beta}  ) ) ( {| <  J_f || {\cal
{\hat T}}_J^{el}  || J_i > | }^2 + {| <  J_f || {\cal {\hat
T}}_J^{mag} || J_i > | }^2
) \\
\nonumber & &  \pm {\mathop\Sigma_{J = 1}} {\hat q} \cdot ({\hat
\nu} - {\vec \beta} )  2 Re [ <  J_f || {\cal {\hat T}}_J^{mag} ||
J_i > {<  J_f || {\cal {\hat T}}_J^{el} || J_i > }^* ]\bigr]~, \Ee
where $(\pm)$ means the case of $\nu ({\bar \nu})$, respectively.
${\vec \nu}$ and $ {\vec k}$ are incident and final lepton
3-momenta, and ${\vec q} = {\vec k} - {\vec \nu}$, ${\vec \beta} =
{\vec k} / \epsilon $ with the final lepton's energy $\epsilon$.
Of course, the extremely relativistic limit (ERL) may yield more
simple formula, but we use the general expression for the
$\nu_{\mu} - A $ reaction. For CC reaction we multiplied Cabbibo
angle $cos ^2 \theta_c$ and include the Coulomb distortion of the
outgoing leptons due to the residual nucleus
\cite{Suzuki06,Ring08}.

Since the Fermi function, $F (Z \pm 1 , \epsilon_f)$, is deduced
for the outgoing electron in the $\beta$ decay \cite{Will74},
which usually assumed s-wave electron, one needs more deliberate
approach for $\nu -A$ reactions, in which outgoing lepton energy
is higher rather than that of the $\beta$ decay. To exactly
describe the Coulomb distortion, one needs to solve the Dirac
equation of the outgoing lepton under the Coulomb potential due to
the residual interactions with daughter nucleus
\cite{kim96,kim01}. But, since the exact solution needs a time
consuming computation, one usually exploits the averaged momentum
of the outgoing particles. Typical method is the effective
momentum approach (EMA), which is able to reproduce exact
calculations more or less \cite{Ring08,Co05-2,Co06}. Here we use
both approaches, the Fermi function and the EMA approach. By
following the prescriptions on Ref.\cite{Ring08,Volpe00}, below
the energy on which both approaches predict same values, the Fermi
function is used, while the EMA is adopted above the energy
region.

\section{Results}
\subsection{$ \nu ({\bar \nu})$ - $^{12}$C reaction}

In Fig.1, we show the cross sections of $^{12}$C$ ( \nu_e , e^-)
^{12}$N$^{*}$ reaction for $J^{\pi} = 0^{\pm} \sim 4 ^{\pm}$
states with dominant transitions ($0^+, 1^{\pm}$ and $ 2^{-}$
states' contributions) and total sum of all states. Other states'
contributions not indicated here are within only a few percentage.
Experimental LSND data for the exclusive $^{12}$C$ ( \nu_e , e^-)
^{12}$N$_{ g.s.(1^+)}$ reaction are taken from Ref.
\cite{lsndedar}. Our results for the Gamow Teller $1^+$ state
clearly reproduce the data. Other $J^{\pi}$ states' contributions
including spin dipole resonances turned out to become also
important for understanding the $\nu - ^{12}$C reaction via CC
\cite{Suzuki06,Ring08}, in specific, beyond 45 MeV energy region.

Results for the NC reaction, $ ^{12}$C$ (\nu , \nu^{'})
^{12}$C$^*$, are shown in Fig.2. Remarkable point is that the GT
transition dominates overwhelmingly the cross section by NC. Below
the 55 MeV region, other transitions contribute within only a few
\%. This is contrast to those for CC reactions. Energy folded (or
flux averaged) cross sections by Eq.(2) for the exclusive and the
inclusive reaction via CC and NC reactions are tabulated in table
1 with other theoretical results and experimental data.

In order to pin down the ambiguities from the nuclear structure,
results for $\beta$ decay are also shown in table I. Here we used
$g_A = 1.23 $ from Ref. \cite{Con72}, which reproduces half lives
of ${\beta}^{-}$ and ${\beta}^{+} (EC)$ decays, as shown in table
1. For pairing interactions, $g_{nn} = 1.6149,~ g_{pp} = 1.4988$
and $g_{np} = 2.0698$ are adjusted to fit the empirical pairing
gaps $\Delta_{nn} = 4.548,~ \Delta_{pp} = 4.430$ and $\delta_{np}
= 2.489$ MeV, respectively \cite{Ch93}.

Our QRPA results reproduce most of the experimental data and show
reasonable consistency with other theoretical model calculations.
But the $ < \sigma >$ for the exclusive reaction, $^{12}$C$( \nu_e
, e^-) ^{12}$N$_{g.s.}$, in table 1 overestimates the data about
10 \% maximally. It might be understood from the following
discussion about the Coulomb distortion.

Main contributions to the folded cross section of the exclusive
reaction come from the energy region $E_{{\nu}_e} \le$ 52 MeV
because of the Michel spectrum, Eq.(3). We divide the energy
region into two regions for the Coulomb correction. In the energy
region below 45 MeV, we take the Fermi function. But the Gamow
Teller $1^+$ state transition, $^{12}$C$( \nu_e , e^-)
^{12}$N$_{g.s.}$, overestimates the experimental data about 7
$\sim$ 8 \% as shown in Fig.1. This is the reason for the
overestimation of the folded cross section.

Here we discuss the Coulomb distortion effect. In Fig.3, we show
three different results ; Fermi function correction, effective
momentum approach (EMA) \cite{Ring08} and no Coulomb correction.
The Fermi function correction turns out to be larger than the EMA
correction in the energy region above 60 Mev, which confirmed
discussions in Refs. \cite{Kolbe03-a,Ring08}. But, in the energy
region $45 \le E_{\nu} \le 60$ MeV, results by both approaches are
nearly indiscernible and well reproduce the experimental data.
However, below 45 MeV, the Fermi function and the EMA overestimate
the cross sections about 7 $\sim$ 8 \% and 10 $\sim$ 12 \%,
respectively, while the case of no Coulomb correction seems to
match with the data. Therefore, for more exact result within 1 or
2 \% for the flux averaged cross section for the $^{12}$C$ ( \nu_e
, e^-) ^{12}$N$_{g.s.}$, more careful treatment of Coulomb
corrections in the low energy region would be necessary.

Our QRPA includes not only proton-proton and neutron-neutron
pairing but also neutron-proton (np) pairing, but the contribution
by the np pairing turns out to be only within 1 $\sim$ 2 \% for
the relevant weak interaction in $^{12}$C, such as $\beta^{\pm}$
decay and the $\nu - ^{12} $C reaction. Since the energy gap
between neutron and proton space is relatively large in light
nuclei, the neutron-proton pairing is expected to be small. But in
the medium-heavy or heavy nuclei, the effect could be larger as
shown in the results for $\nu-^{56}$Ni$(^{56}$Fe) reactions.

\subsection{$ \nu ({\bar \nu})$ - $^{56}$Ni and
$ \nu ({\bar \nu})$ - $^{56}$Fe reaction}

Here we calculate reactions of $^{56}$Ni$ ( \nu_e , \nu_e^{'})
^{56}$Ni$^{*}$, $^{56}$Ni$ ( {\bar \nu}_e , {\bar \nu}_e^{'})
^{56}$Ni$^{*}$ via NC, and $^{56}$Fe$ ( \nu_e , e^{-})
^{56}$Co$^{*}$ via CC. Results for $\nu ({\bar \nu}_e) - ^{56}$Ni
by NC are shown in Fig.4 and 5. Remarkable difference between
$\nu_e - $ and ${\bar \nu}_e - ^{56} Ni$ is the magnitude of cross
sections, in which cross sections by $\nu_e$ reaction are about 40
\% larger than those by ${\bar \nu}_e$ reaction. But the cross
section is dominated by the GT transition. Contributions by higher
multipole transitions are shown to be within only a few \%,
similarly to the case of $^{12}$C$ ( \nu_e , \nu_e^{'} )
^{12}$C$^*$. Higher multipole contributions to the NC reaction
turns out to be much smaller than those to the CC reaction. It is
also confirmed in $^{56}$Fe$ ( \nu_e , e^{-}) ^{56}$Co$^{*}$
reaction as shown in Fig.6.

Fermi function for the Coulomb corrections is used on the energy
region below 45 MeV and the EMA is taken beyond 45 MeV. For
pairing interactions, $g_{nn} = 1.3134 (1.0244), g_{pp} =
1.1558(1.0340)$ and $g_{np} = 1.2785(1.3057)$ are adjusted to the
empirical pairing gaps $\Delta_{nn} = 2.148(1.425), \Delta_{pp} =
2.078(1.572)$ and $\delta_{np} = 1.102(0.336)$ MeV for $^{56}$Ni
$(^{56}$Fe) nucleus, respectively \cite{Ch93}.

Since we have an experimental data for $^{56}$Fe$ ( \nu_e , e^{-})
^{56}$Co$^{*}$ by the DAR neutrino given as energy weighted cross
section, which is the only experimental $\nu - A$ reaction data
beyond $^{12}C$, we compared our results to the experimental data.
If the np pairing is switched off, we obtain $< \sigma>$ = 141.8
$\times 10^{-42}  cm^{2}$. But the np pairing enhanced it to 173.5
$\times 10^{-42}  cm^{2}$ which is located within the experimental
data 256$\pm$108$\pm$43 $\times 10^{-42}  cm^{2}$. As discussed in
Ref. \cite{Ring08}, previous QRPA calculations without np pairing
could not reproduce the experimental data.

In Fig.7, we show GT strength feasible from the (p,n) or (n,p)
reaction on $^{56}$Fe or $^{56}$Ni target. They are calculated as
\begin{equation}
B(GT_{\pm}) = { 1 \over  {2 J_i + 1  }} {| < f ||
{\mathop\Sigma_{k}  \sigma_k \tau_{k_{\mp}}} || i >|}^2 ~.
\end{equation}
Results in Fig.7 did not use the quenching factor. Our results for
total GT strength with the quenching factor $f_q= 0.74$ are 11.38
and 4.41 for $B(GT_{\mp})$. They are consistent with those of
experimental data, $9.9 \pm 2.4$ \cite{Ra83} and $2.8 \pm 0.3$
\cite{Ca99}, respectively. As shown in Fig.7, contributions from
the higher excited states lead to a bit larger value for
$B(GT_{+})$ than the data. Similarly to the results by the shell
model \cite{Suzuki09}, which is a hybrid model {\it i.e.} shell
model for GT and IAS transition and QRPA model for high multipole
transitions, our results reproduced the experimental data by using
the same quenching factor as the shell model. We expect that more
fruitful data for the relevant weak transitions, which enables to
constraint the ambiguities from the nuclear structure.

For higher multipoles we did not use the quenching factor by
following the discussions at Ref. \cite{Suzuki09}. But larger
contributions compared to those by Ref. \cite{Suzuki09} are
obtained for the higher multiople transitions. In Ref.
\cite{Suzuki09}, 80 \% of the data is explained by the GT and IAS
transition. Therefore our results explain systematically the $\nu
- ^{56}$Fe and $\nu- ^{56}$Ni reaction as well as the GT strength
in the single $\beta$ decay in a systematically organized QRPA
model. For the supernovae application, we show the cross sections
averaged by the SN neutrino spectrum in Eq.(1) in Fig. 8. Cross
section via CC are much larger, about 3 times, than those by NC,
irrespective of nucleus species. Application to the $\nu$ process
for other relevant nuclei are in progress.

\section{Summary and Conclusion}

We applied the QRPA to the $\nu - A$ reaction by including
multipole transitions up to $J^{\pi} = 4^{\pm}$ with explicit
momentum dependence and evaluated the $\nu$-reaction to $^{12}$C,
$^{56}$Ni and $^{56}$Fe target. Our results for the $\nu ({\bar
\nu}) - A$ reaction show quite consistent results with available
experimental data and other theoretical calculations.

Remarkable points from our results are summarized as follows.
Firstly, not only GT, IAS and spin dipole transitions but also
other multipole transitions could play roles for $\nu - A$
reaction via CC below the quasi elastic peak. But NC reaction is
dominated by the GT transition.

Secondly, np pairing as well as nn and pp pairing correlations act
as important ingredients for describing the BCS ground state, and
contribute to some extent to the cross section of the $\nu - A$
reaction. In particular, it affects the results for medium heavy
nuclei such as $^{56}$Ni and $^{56}$Fe, which was already expected
from our previous results for $2 \nu 2 \beta$ and 0$\nu 2 \beta$
decays.

Finally, for more exact evaluation within a few \% for the flux
averaged cross section of the future would-be experimental data,
more deliberately chosen Coulomb correction would be necessary. In
particular, experimental data via NC would be desirable to pin
down the ambiguities on the Coulomb corrections.

The QRPA is a very efficient method to consider multi-particle and
multi-hole interactions and their configuration mixing, which
successfully described nuclear reactions sensitive on the nuclear
structure, such as $2 \nu 2 \beta$ and 0$\nu 2 \beta$ decays.
Therefore the ambiguities feasible due to the nuclear structure
can be pinned down by reproducing the data related to the $\beta$
and $\beta \beta$ decay. The extension of our QRPA calculation for
$\nu - A$ reaction to explicitly include the deformation
\cite{Si03}, which turns out to be so important for the exotic
nuclei frequently appeared, is under progress. It enables us to
perform various nuclear weak reactions for stable and unstable
nuclei.

This work was supported by the Korea Research Foundation Grant
funded by the Korean Government(MOEHRD, Basic Research Promotion
Fund)(KRF-2006-331-C00078) and one of author, Cheoun, was
supported by the Soongsil University Research Fund.

\newpage

\vspace{5mm}
\begin{table}
\caption[bb]{Comparison of calculated and measured flux averaged
cross sections for the $\nu - ^{12}$C reaction in units of
$10^{-42} cm^2$, and half life time of neighboring nuclei. The
cross sections are folded by the corresponding DAR neutrino
spectra, where the Michel spectrum is used for $\nu_e$ and
$\nu_{\mu}$ energy is fixed at 29.8 MeV. "K" and "L" mean Karmen
and LSND groups results, respectively. Shell Model(SM) and
Continuum RPA(CRPA) results are cited from Ref.\cite{Suzuki06} and
Ref.\cite{Kolbe03-a}, respectively. ($9.834^*$) is a result with
no Coulomb correction.}
 \setlength{\tabcolsep}{3.0 mm}
\begin{tabular}{cccc}\hline
                            & $^{12}$C$ ( \nu_e , e^-) ^{12}$N$_{g.s.}$  &
          $^{12}$C$( \nu_e , e^-) ^{12}$N$^*$   & $\beta^{(-)} : (^{12}$B$(1^+) \rightarrow ^{12}$C)      \\
                                 \hline\hline
  Exp.   & ${8.9 \pm 0.3 \pm 0.9 }$~{\cite{lsnd}"L"}  & ${4.3 \pm 0.4 \pm 0.6}$~{\cite{lsnd}"L"}
     & 23.6 ms \\
     & ${9.1 \pm 0.5 \pm 0.8}$~{\cite{karmenedar}"K"} & ${5.1 \pm 0.6 \pm 0.5}$~{\cite{karmencc}"K"} &
     \\ \hline
  Ours  & 11.53 $({9.834}^*$)& 6.1 & 21.33 ms \\ \hline
  SM & 9.06 $\sim$ 8.48 &  5.22 $\sim$ 4.87  &  \\ \hline
  CRPA  & 8.9  &  5.4 & \\ \hline\hline
         &           $^{12}$C$ [( \nu_e , \nu_e^{'}) + ( {\bar \nu}_{\mu} ,
         {\bar \nu}_{\mu}^{'})] ^{12}$N$_{g.s.}$
           &  $^{12}$C$ [( { \nu}_{\mu} , { \nu}_{\mu}^{'})] ^{12}$C$^*$
           &  $\beta^{(+)} (EC) : (^{12}$N$(1^+) \rightarrow ^{12}$C)  \\
                                 \hline\hline
  Exp.       & ${10.4 \pm 1.0 \pm 0.9}$~{\cite{karmenedar}"K"}
  & ${3.2 \pm 0.5 \pm 0.4}$~{\cite{Armb98}"K"} & 11.0 ms\\ \hline
  Ours & 9.92  &  3.60   & 10.34 ms \\ \hline
  SM  & 9.76 $\sim$ 8.27    & 2.68 $\sim$ 2.26 &   \\ \hline
  CRPA & 10.5     &   &  \\ \hline

\end{tabular}
\label{tab:result1}
\end{table}

\newpage

\begin{figure}
\includegraphics[width=0.85\linewidth]{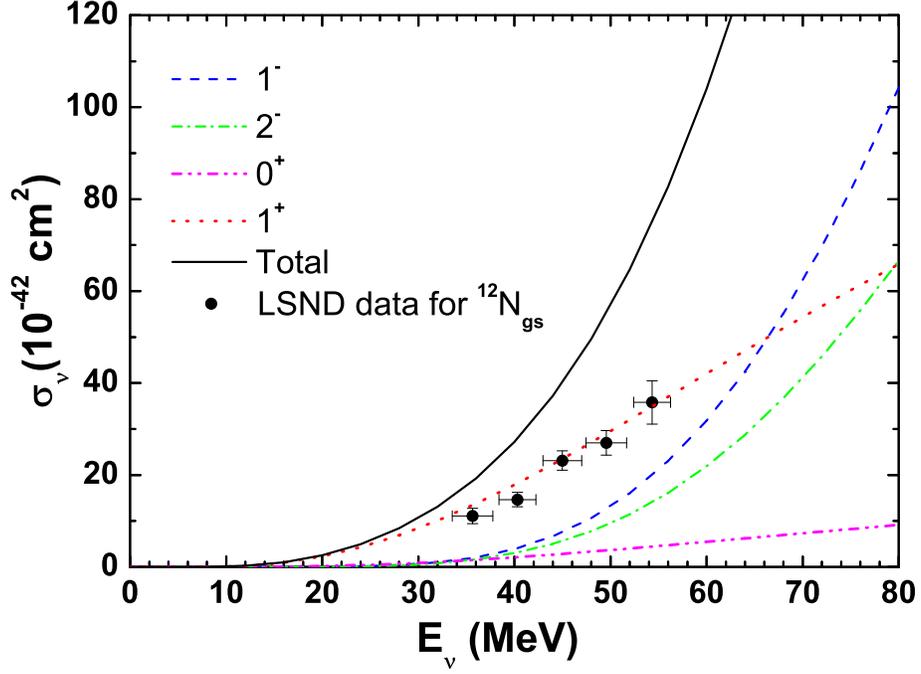}
\caption{(Color online) Cross sections of the $^{12}$C$ ( \nu_e ,
e^-) ^{12}$N$^{*}$ reaction for $J_{\pi} = 0^{\pm} \sim 4^{\pm}$
states. Main multipole states ($0^+ , 1^{\pm}$ and $2^{-}$)
contributions and total sum are presented. Cross section via
$J^{\pi} = 1^+ $ state is compared to the LSND experimental data,
$^{12}$C$ ( \nu_e , e^-) ^{12}$N$_{g.s. ( 1^+)}$ \cite{lsndedar}.}
\label{fig1}
\end{figure}

\begin{figure}
\includegraphics[width=0.85\linewidth]{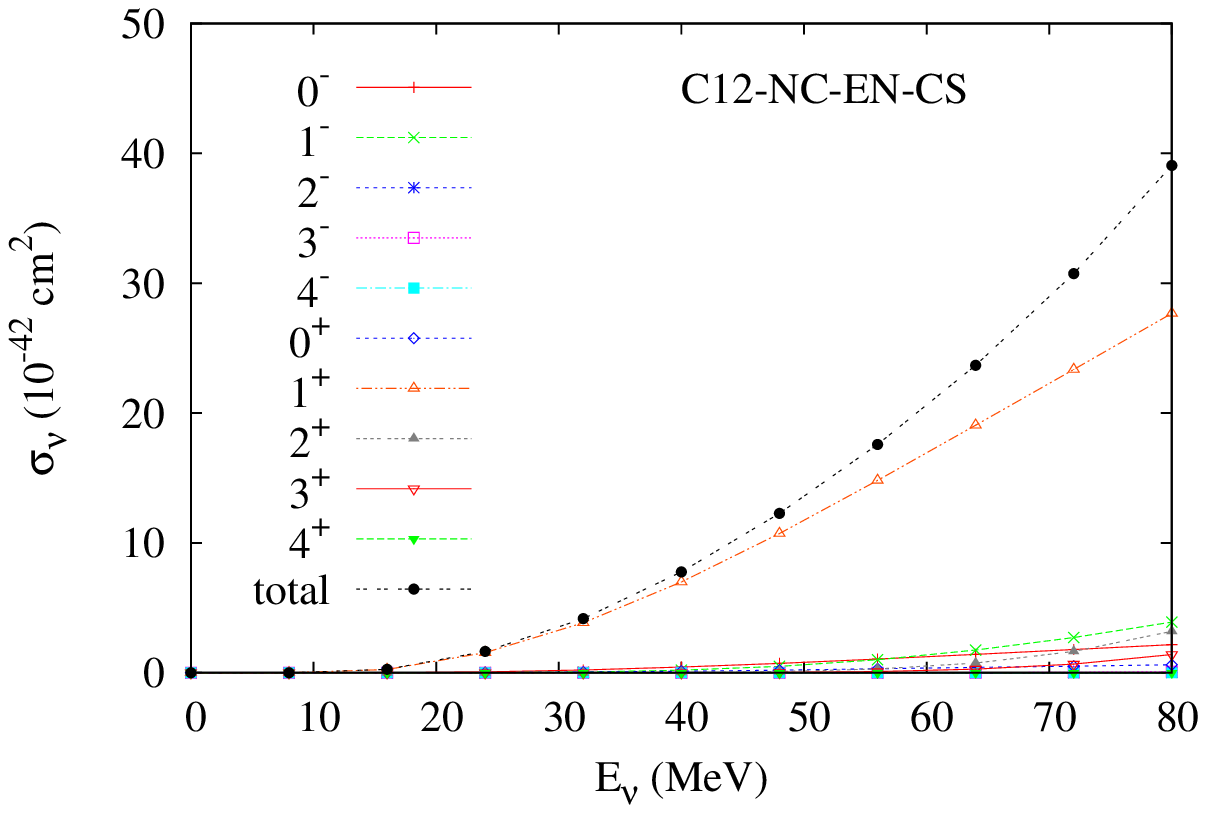}
\caption{(Color online) Cross sections of the $^{12}$C$ ( \nu_e ,
\nu_e^{'}) ^{12}$C$^{*}$ reaction for $J_{\pi} = 0^{\pm} \sim
4^{\pm}$ states. Each multipole state contribution and total sum
are presented.} \label{fig2}
\end{figure}

\begin{figure}
\includegraphics[width=0.85\linewidth]{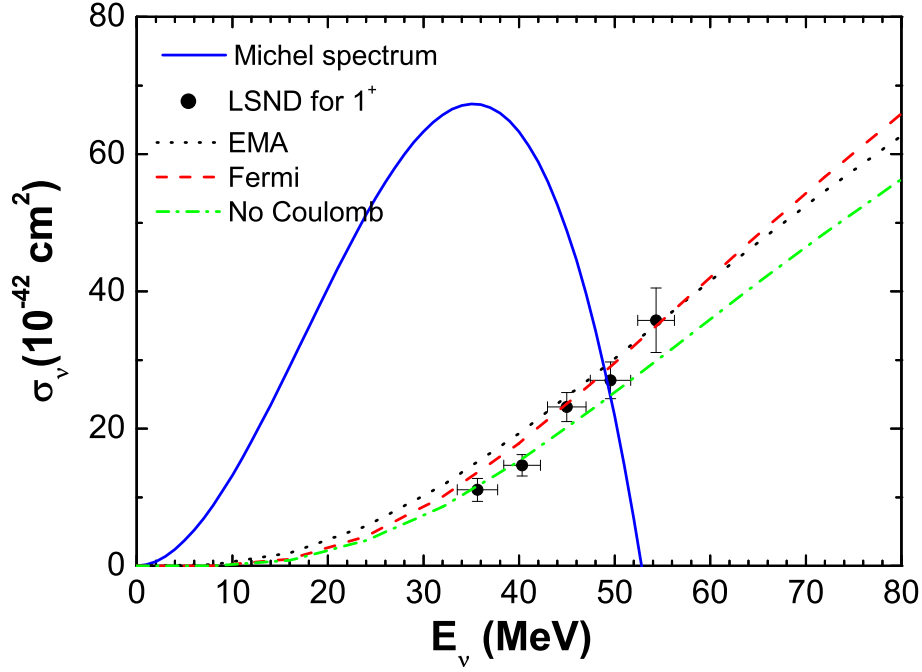}
\caption{(Color online) Comparison of two different Coulomb
corrections and no Coulomb correction for the $^{12}$C$ ( \nu_e ,
e^-) ^{12}$N$_{g.s. ( 1^+)}$ reaction. Arbitrary normalized Michel
spectrum for DAR neutrino is also presented to indicate the
contribution by the incident neutrino energy bin to the flux
averaged cross sections. Data point for the $J^{\pi} = 1^+ $ state
is the same as Fig.1 {\it i.e.} the LSND data for $^{12}$C$ (
\nu_e , e^-) ^{12}$N$_{g.s. ( 1^+)}$ \cite{lsndedar}.}
\label{fig3}
\end{figure}


\newpage

\begin{figure}
\includegraphics[width=0.85\linewidth]{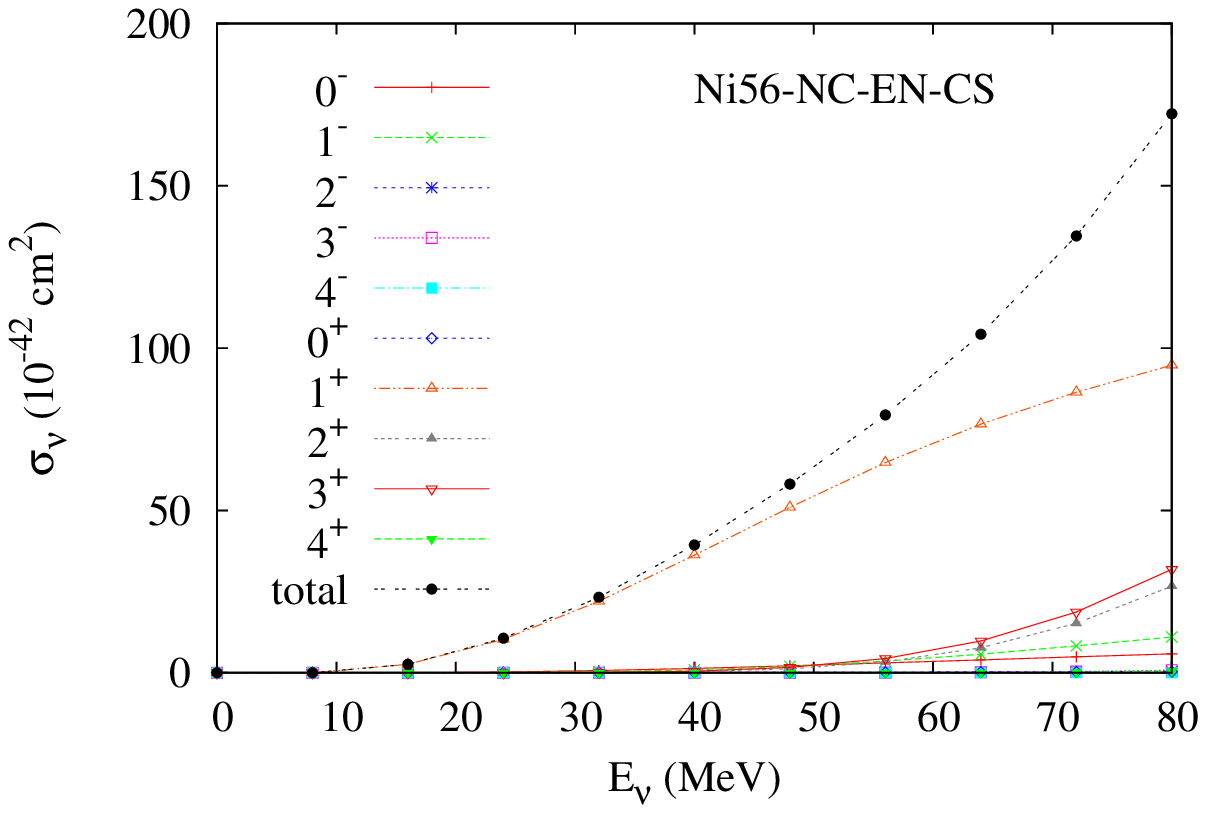}
\caption{(Color online) Cross sections of the $^{56}$Ni$ ( \nu_e ,
\nu_e^{'} ) ^{56}$Ni$^{*}$ reaction for $J_{\pi} = 0^{\pm} \sim
4^{\pm}$ states. Each multipole state contribution and total sum
are presented.} \label{fig4}
\end{figure}

\newpage

\begin{figure}
\includegraphics[width=0.85\linewidth]{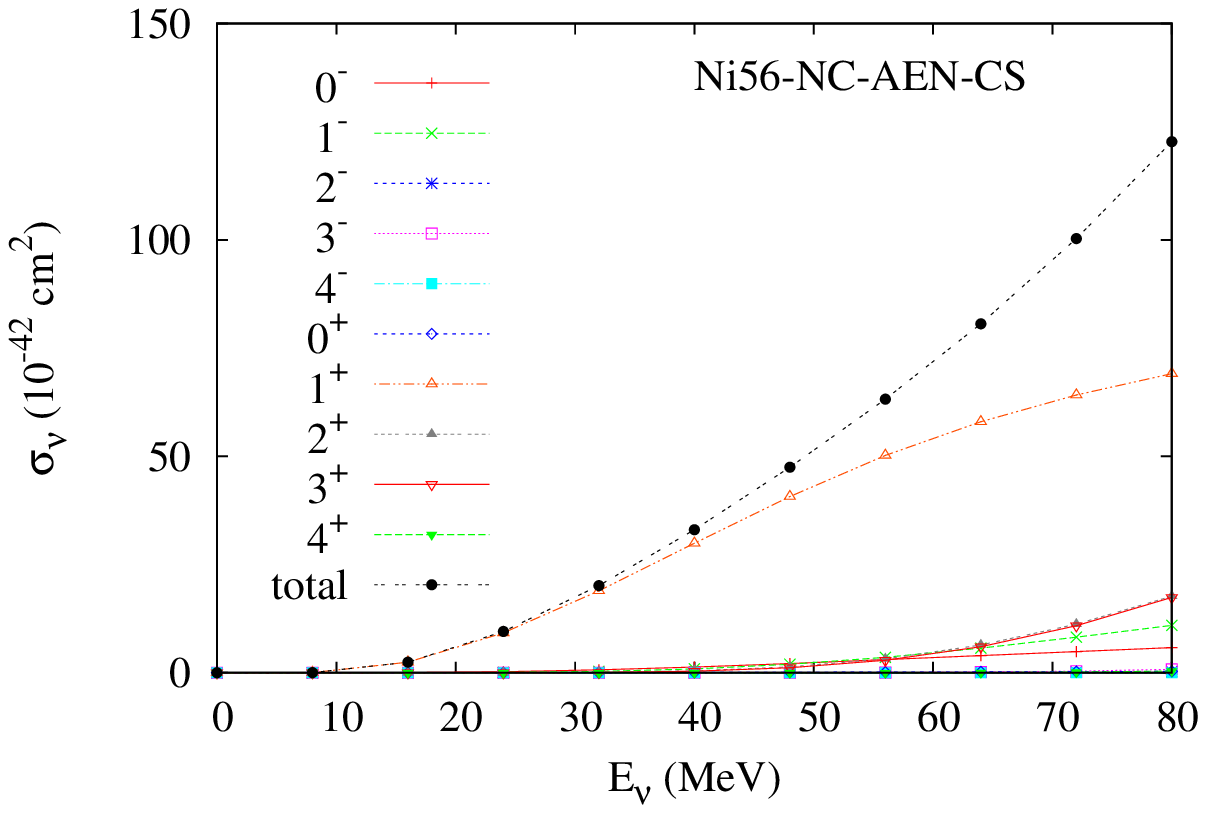}
\caption{(Color online) Cross sections of the $^{56}$Ni$ ( {\bar
\nu}_e , {\bar \nu}_e^{'} ) ^{56}$Ni$^{*}$ reaction for $J_{\pi} =
0^{\pm} \sim 4^{\pm}$ states. Each multipole state contribution
and total sum are presented.} \label{fig5}
\end{figure}


\begin{figure}
\includegraphics[width=0.85\linewidth]{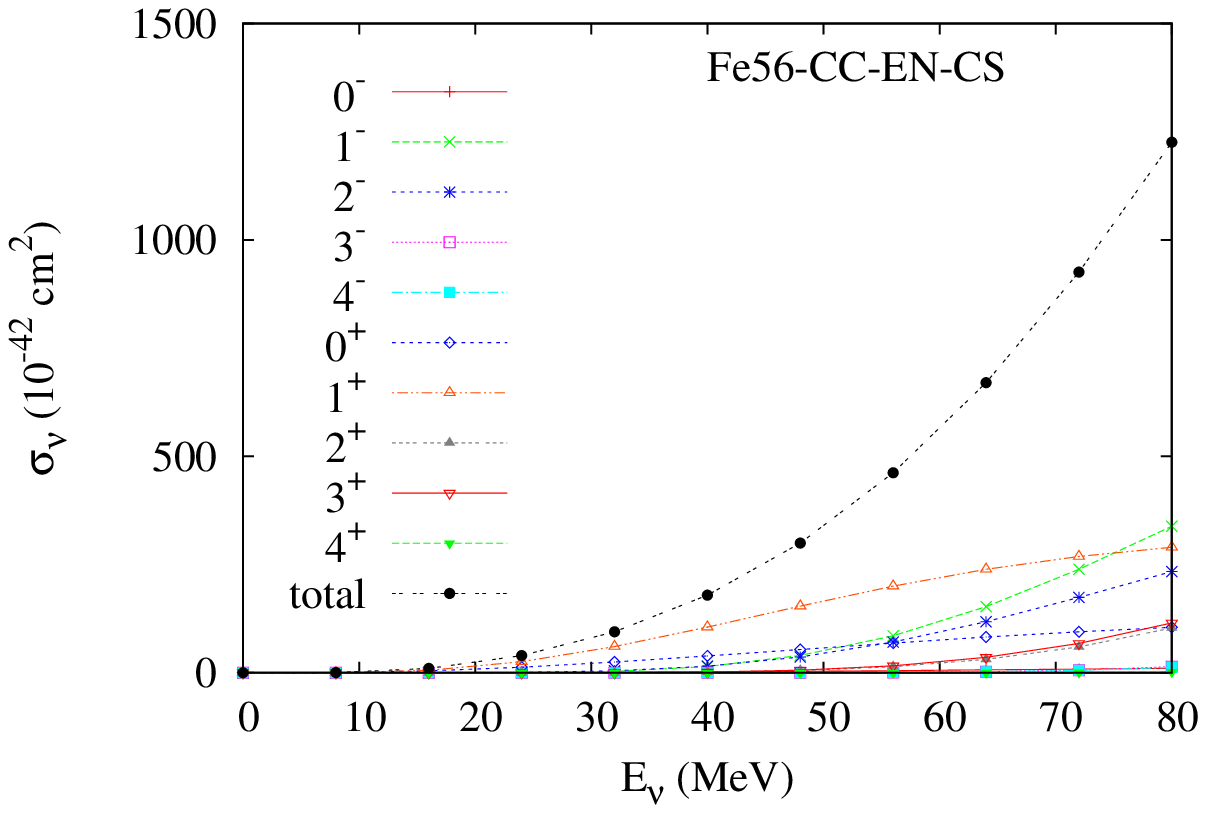}
\caption{(Color online) Cross sections of the $^{56}$Fe$ ( \nu_e ,
e^{-} ) ^{56}$Co$^{*}$ reaction for $J_{\pi} = 0^{\pm} \sim
4^{\pm}$ states. Each multipole state contribution and total sum
are presented.} \label{fig6}
\end{figure}

\begin{figure}
\centering
\includegraphics[width=7.5cm]{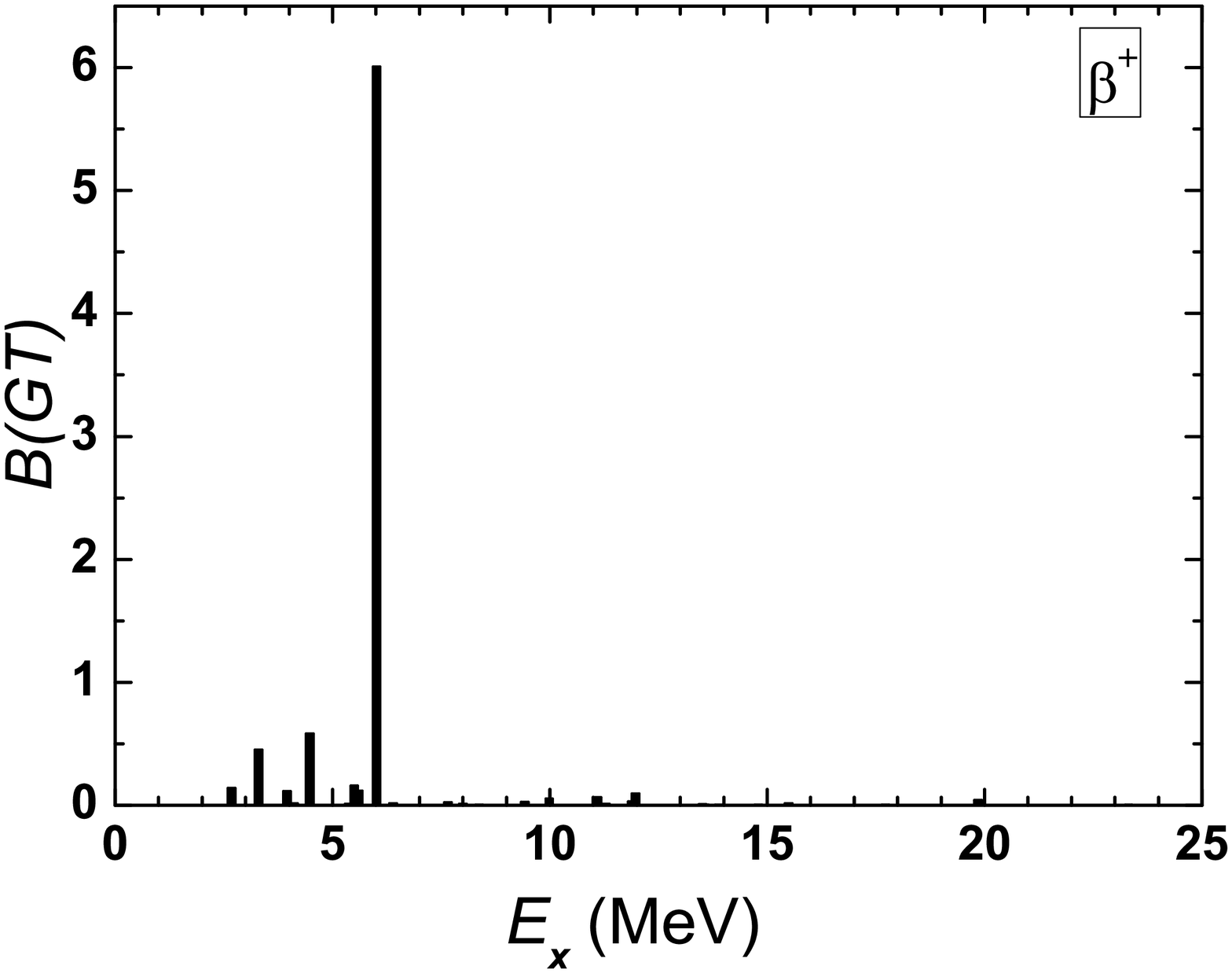}
\includegraphics[width=7.5cm]{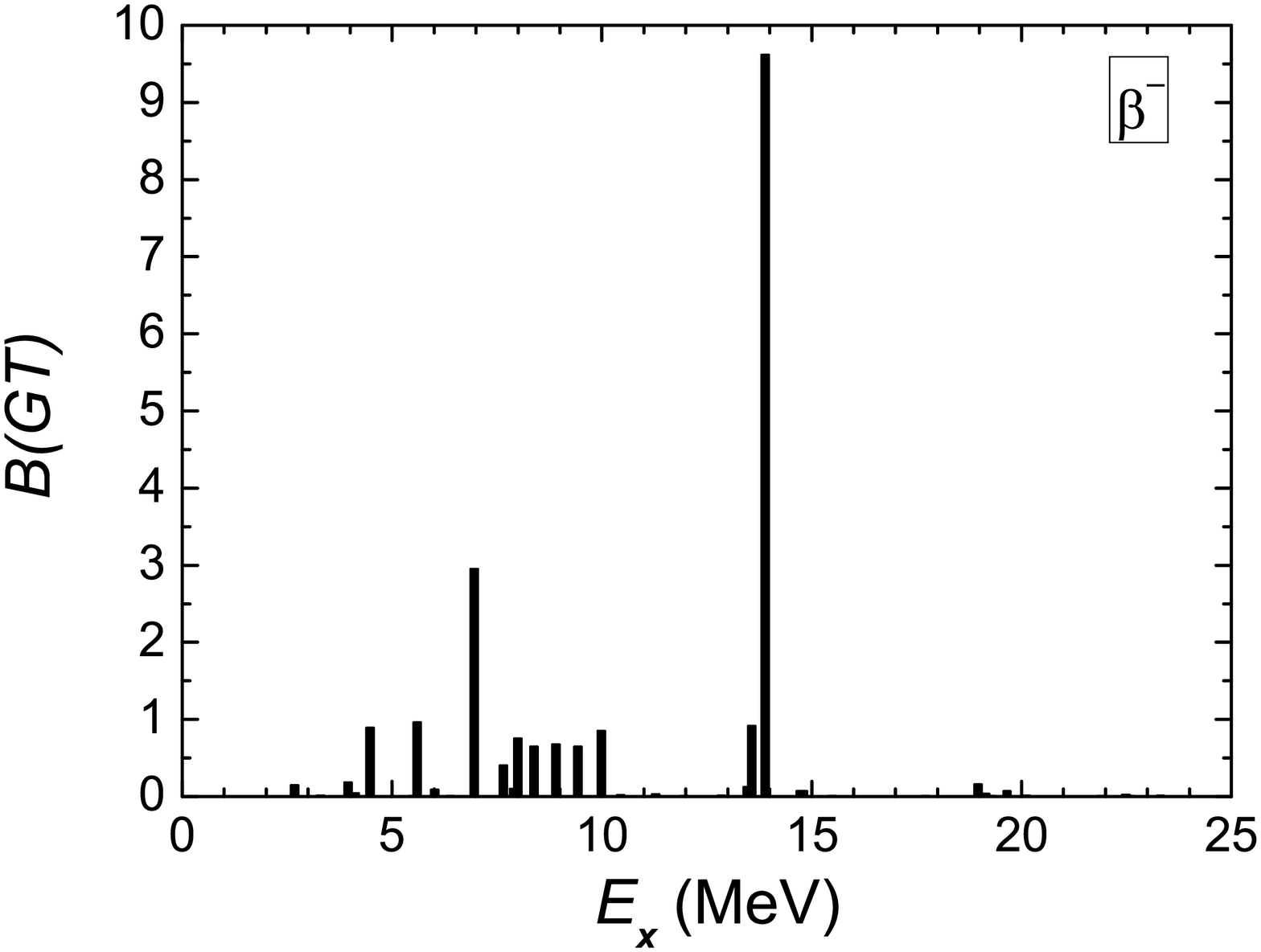}
\caption{(Color online) Gamow-Teller strength $B(GT_{\pm})$ for
$^{56}$Fe} \label{fig7}
\end{figure}

\begin{figure}
\includegraphics[width=0.85\linewidth]{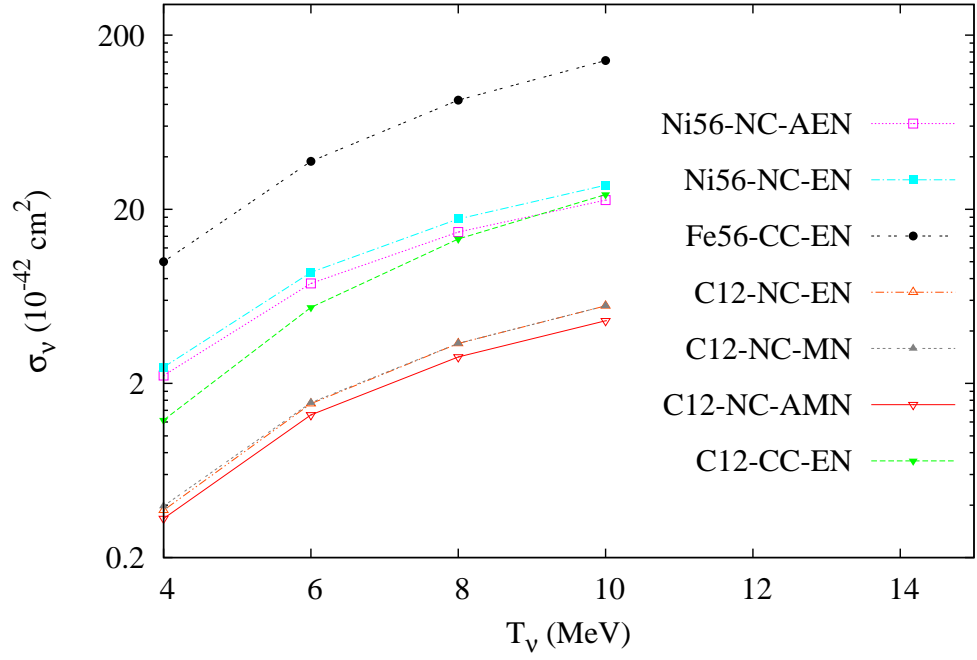}
\caption{(Color online) Temperature dependence of the energy
weighted cross section, Eq. [2] for $\nu - ^{12}$C, -$^{56}$Ni and
-$^{56}$Fe reactions, where neutrino spectrum for the SN, Eq.(1),
is exploited.} \label{fig8}
\end{figure}


\begin{thebibliography}{140}
\par
\vskip0.5cm
\par
\def\pr{Phys. Rev.}
\def\prl{Phys. Rev. Lett.}
\def\nc{Nucl. Phys.}
\def\pl{Phys. Lett.}
\def\nuc{Nuovo. Cim.}
\def\pro{Prog. Theo. Phys}
\def\so{Sov.J.Nucl.Phys.}
\def\can{Can. J. Phys.}
\par
\bibitem{Kl92} W. L. Freedman and M. S. Turner, Rev. of Modern Phys. {\bf 75},
1433 (2003).

\bibitem{Con72} J. S. O'Connell, T. W. Donnelly, and J. D.
Walecka, Phys. Rev. C {\bf 6}, 719 (1972).

\bibitem{Suzuki06} T. Suzuki, S. Chiba, T. Yoshida, T. Kajino, T.
Otsuka, Phys. Rev. C {\bf 74}, 034307 (2006).

\bibitem{Woosley90} S. E. Woosley, D. H. Hartmann, R. D. Hoffmann,
and W. C. Haxton, Astrophys. J. {\bf 356}, 272 (1990).


\bibitem{yoshida08} T. Yoshida, T. Suzuki, S, Chiba, T. Kajino, H.
Yokomukura, K. Kimura, A. Takamura, H. Hartmann, Astro. Phys. J.
{\bf 686}, 448 (2008).
\bibitem{Suzuki09} T. Suzuki, M. Honma, K. Higashiyama, T. Yoshida, T. Kajino, T.
Otsuka, H. Umeda, and K. Nomoto, Phys. Rev. C {\bf 79}, 061603
(2009).

\bibitem{Heg} A. Heger, E. Kolbe, W.C. Haxton, K. Langanke,
G. Martinez-Pinedo and S.E. Woosley, Phys. Lett. B {\bf 606}, 258
(2005).
\bibitem{Wana06} Shinya Wanajo, Astrophys. J. {\bf 647}, 1323 (2006).
\bibitem{Kolbe03-a} E. Kolbe, K. Langanke, G Martinez-Pinedo and P. Vogel, J. Phys.
G {\bf 29}, 2569 (2003).
\bibitem{Ring08} N. Paar, D. Vretenar, T. Marketin, and P. Ring,
Phys. Rev. C {\bf 77}, 024608 (2008).

\bibitem{lsndedar} C. Athanasopoulos {\it et
al.}, (LSND Collaboration),  Phys. Rev. C {\bf 55}, 2078 (1997).
\bibitem{lsnd} L. B. Auerbach {\it et al.} (LSND Collaboration),
Phys. Rev. C {\bf 64}, 065501 (2001).


\bibitem{karmenedar} B. E. Bodmann {\it et al.}, (KARMEN
Collaboration), Phys. Lett. {\bf B 332}, 251 (1994).

\bibitem{karmencc} R. Maschuw, Prog. Part. Nucl. Phys. {\bf 40}, 183 (1998).
\bibitem{Armb98} B. A. Armbruster {\it et al.} (KARMEN
Collaboration), Phys. Lett. {\bf B 423}, 15 (1998).


\bibitem{zucch02}P. Zucchelli, Phys. Lett. B {\bf 532}, 166
(2002).

\bibitem{volpe}Cristina Volpe, J. Phys. G {\bf 34}, R1 (2007).


\bibitem{sato}Joe Sato, Phys. Rev. Lett. {\bf 95}, 131804 (2005).

\bibitem{bernabeu}Jose Bernabeu, Jordi Burguet-Castell, Catalina
Espinoza, and Mats Lindroos, J. Hep. {\bf 12} 014 (2005).


\bibitem{Haus52} W. Hauser and H. Feshbach, Phys. Rev. {\bf 87},
366 (1952).


\bibitem{Volpe00} C. Volpe, N. Auerbach, G. Colo, T. Suzuki, and
N. Van Giai, Phys. Rev. C {\bf 62}, 015501 (2000).




%


\bibitem{Kolbe95}E. Kolbe, K. Langanke, F.-K.Thielemann, and P.
Vogel, Phys. Rev. C {\bf 52}, 3437 (1995).



\bibitem{Kolbe03}E. Kolbe, Nucl. Phys. {\bf A719}, 135c (2003).

\bibitem{Ch93} M. K. Cheoun, A. Bobyk, Amand Faessler, F. Simcovic and
G. Teneva, {\nc} {\bf {A561}}, 74 (1993) ; {\nc} {\bf {A564}}, 329
(1993); M. K. Cheoun, G. Teneva and Amand Faessler, Prog. Part.
Nuc. Phys. {\bf 32}, 315 (1994) ; M. K. Cheoun, G. Teneva and
Amand Faessler, {\nc} {\bf A587}, 301 (1995).


\bibitem{Ryck02} N. Jachowicz, K. Heyde, J.
Ryckebusch, and S. Rombouts, Phys. Rev. C {\bf 65}, 025501 (2002).
\bibitem{Co05-1} Antonio Bortrungo, Giampaolo Co', Nucl. Phys. {\bf
A761}, 200 (2005).


\bibitem{Ba63} M. Baranger, {\pr}{\bf 130}, 1244 (1963).


\bibitem{Ho81} K. Holinde, Phys. Rep. {\bf 68}, 121 (1981).

\bibitem{Don79} T. W. Donnelly and W. C. Haxton, ATOMIC DATA AND
NUCLEAR DATA {\bf 23}, 103 (1979).

\bibitem{Wal75} J. D. Walecka, in {\it Muon Physics}, edited by V.
H. Huges and C. S. Wu (Academic, New York, 1975), Vol II.

\bibitem{ch08} Myung-Ki Cheoun and K. S. Kim, J. Phys. {\bf G 35}
065107 (2008).




\bibitem{giusti1}Andrea Meucci, Carlotta Giusti, and Franco Davide
Pacati, Nucl. Phys. {\bf A739}, 277 (2004); Nucl. Phys. {\bf
A744}, 307 (2004); Nucl. Phys. {\bf A773}, 250 (2006).

\bibitem{musolf}M. J. Musolf and T. W. Donnelly, Nucl. Phys. {\bf
A546}, 509 (1992).








\bibitem{Will74} D. H. Wilkinson and B. E. F. Macefield, Nucl.
Phys. {\bf A 232}, 58 (1974).


\bibitem{kim96} K. S. Kim, L. E. Wright, Yanhe Jin, and D. W.
Kosik, Phys. Rev. {\bf C 54}, 2515 (1996).
\bibitem{kim01} K. S. Kim, L. E. Wright, and D. A. Resler, Phys. Rev. {\bf C 64}, 044607 (2001).

\bibitem{Co05-2} A. Bortrungo and G. Co', Eur. Phys. J. {\bf
A24S1}, 109, (2005).
\bibitem{Co06} Giampaolo Co', Acta Physica Polonica B {\bf
37}, 2235, (2006).

\bibitem{Ra83} J. Rapaport {\it et. al}, Nucl. Phys. {\bf A 410},
371 (1983).
\bibitem{Ca99} E. Caurier, K. Langanke, G. Martinez-Pinedo, and F.
Nowacki, Nucl. Phys. {\bf A 653}, 439 (1999).

\bibitem{Si03} F. Simkovic, L. Pacearescu, and A. Faessler, {\pr} {\bf C68}, 054319 (2003);
{\nc} {\bf A733}, 321 (2004).





\end{thebibliography}
\end{document}